\documentclass[runningheads]{llncs}

\usepackage[height=235mm,width=155mm,center]{crop}
\usepackage{graphicx}

\usepackage{textpos}

\usepackage[utf8]{inputenc}

\usepackage{marvosym}
\usepackage{./orcidlink}

\renewcommand{\orcidID}[1]{\footnotesize\orcidlink{#1}\normalsize}

\usepackage{graphicx}
\usepackage[caption=false]{subfig}
\usepackage{xcolor}

\usepackage{listings}

\usepackage{booktabs}
\usepackage[inline]{enumitem}

\setenumerate{label=(\roman*), leftmargin=2em}

\usepackage{wasysym}

\usepackage[hidelinks,implicit=true,bookmarksdepth=2]{hyperref}

\hypersetup{
	colorlinks = true,
	citecolor=blue,
	linkcolor=blue,
	pdftitle={Evaluation of Risk-based Re-Authentication Methods},
	pdfauthor={Stephan Wiefling,Tanvi Patil,Markus Dürmuth,Luigi Lo Iacono},
	pdfsubject={Risk-based Authentication (RBA) is an adaptive security measure that improves the security of password-based authentication by protecting against credential stuffing, password guessing, or phishing attacks. RBA monitors extra features during login and requests for an additional authentication step if the observed feature values deviate from the usual ones in the login history. In state-of-the-art RBA re-authentication deployments, users receive an email with a numerical code in its body, which must be entered on the online service. Although this procedure has a major impact on RBA's time exposure and usability, these aspects were not studied so far. We introduce two RBA re-authentication variants supplementing the de facto standard with a link-based and another code-based approach. Then, we present the results of a between-group study (N=592) to evaluate these three approaches. Our observations show with significant results that there is potential to speed up the RBA re-authentication process without reducing neither its security properties nor its security perception. The link-based re-authentication via "magic links", however, makes users significantly more anxious than the code-based approaches when perceived for the first time. Our evaluations underline the fact that RBA re-authentication is not a uniform procedure. We summarize our findings and provide recommendations.},
	pdfkeywords={Risk-based Authentication (RBA), Re-authentication, Usable Security}
}

\begin{document}
	\title{Evaluation of Risk-based\\Re-Authentication Methods}
	\author{Stephan Wiefling\inst{1,3}\textsuperscript{\small(\Letter)}\orcidID{0000-0001-7917-6065} \and
		Tanvi Patil\inst{2}\orcidID{0000-0003-3640-1124}%
		\and\\
		Markus Dürmuth\inst{3} \and
		Luigi {Lo Iacono}\inst{1}\orcidID{0000-0002-7863-0622}
	}
	\authorrunning{S. Wiefling et al.}

	\institute{%
		H-BRS University of Applied Sciences, Sankt Augustin, Germany \email{\{stephan.wiefling,luigi.lo\_iacono\}@h-brs.de} \and
		University of North Carolina at Charlotte, Charlotte, NC, USA \email{tpatil@uncc.edu} \and
		Ruhr University Bochum, Bochum, Germany 
		\\ \email{\{stephan.wiefling,markus.duermuth\}@rub.de}
	}
	\maketitle              %
	\begin{textblock}{9}(0.0,8.35)
		\noindent
		\scriptsize Postprint version of a paper accepted for IFIP SEC 2020.\\The final publication is available at Springer via\\ \url{http://dx.doi.org/10.1007/978-3-030-58201-2_19}
	\end{textblock}
	\vspace{-1.3em}
	\begin{abstract}
Risk-based Authentication (RBA) is an adaptive security measure that improves the security of password-based authentication by protecting against credential stuffing, password guessing, or phishing attacks. RBA monitors extra features during login and requests for an additional authentication step if the observed feature values deviate from the usual ones in the login history. In state-of-the-art RBA re-authentication deployments, users receive an email with a numerical code in its body, which must be entered on the online service. Although this procedure has a major impact on RBA's time exposure and usability, these aspects were not studied so far.
We introduce two RBA re-authentication variants supplementing the de facto standard with a link-based and another code-based approach. Then, we present the results of a between-group study (N=592) to evaluate these three approaches. Our observations show with significant results that there is potential to speed up the RBA re-authentication process without reducing neither its security properties nor its security perception. The link-based re-authentication via ``magic links'', however, makes users significantly more anxious than the code-based approaches when perceived for the first time. Our evaluations underline the fact that RBA re-authentication is not a uniform procedure. %
We summarize our findings and provide recommendations. 		
	\end{abstract}
	\vspace{-2em}
\section{Introduction}
Passwords were and continue to be the predominant authentication mechanism of online services~\cite{quermann_state_2018}. However, threats to password-based authentication are increasing, e.g, by large-scale password database leaks and credential stuffing~\cite{thomas_protecting_2019}. %
Therefore, website operators have to provide additional or alternative authentication mechanisms to adequately protect their users. Two-factor authentication (2FA) is one such measure which is widely used but has proven to be unpopular among users~\cite{milka_anatomy_2018}. %
Biometric authentication %
is considered impractical for large-scale online services since it requires special hardware and active participation from the user~\cite{gaddam_usage_2019}.
For these reasons, several large online services deployed risk-based authentication (RBA) to protect their users~\cite{wiefling_is_2019}. RBA is an adaptive authentication measure that provides high security with minimal impact on user interaction, and thus has the potential to be more accepted by users than 2FA. Moreover, RBA is recommended in the NIST digital identity guidelines to mitigate account takeover~\cite{grassi_digital_2017}.

During password entry, RBA monitors additional features, e.g., IP address or user agent, and requests for re-authentication when a particular risk is detected%
~\cite{freeman_who_2016}. In state-of-the-art deployments, the re-authentication is mostly based on email address verification~\cite{wiefling_is_2019}. Here, the user receives an email with a multi-digit code in the email body that has to be entered on the online service. %

Despite its clear presence in RBA deployments, there are, to the best of our knowledge, no studies that evaluate this state-of-the-art re-authentication method%
. Investigating different devices is important for RBA because push notifications from mobile email apps can make it possible to check emails on mobile devices faster than on desktop devices. Furthermore, using the website on a desktop PC and checking email on a mobile device can slow down the re-authentication process since the code has to be typed in manually. %
We also discovered that online services using RBA offer different email verification methods for account registration than for RBA re-authentication. When registering an account, the user received either an email with a digit code in the email subject and body, or a verification link. Thus, we wondered why these verification methods are not being used in the RBA re-authentication context so far and whether they have the potential to improve the RBA experience while maintaining the same level of security. %
To close this gap, we formulated the following research questions.

\subsubsection{Research Questions.}
\label{subsection:research-questions}

With these %
questions, we aim to give answers as to whether the widespread email-based re-authentication method can be improved by other approaches and how all of these methods are perceived by users.

\newlist{RQLIST}{enumerate}{1}
\setlist[RQLIST]{label=\bfseries RQ\arabic*:, leftmargin=3.2em, parsep=0em}

\newlist{RQ2LIST}{enumerate}{2}
\setlist[RQ2LIST]{label=\bgroup\bfseries \alph*)\egroup,leftmargin=1.6em, parsep=0em}

\begin{RQLIST}
	\item \begin{RQ2LIST}
		\item How does link-based re-authentication affect the authentication time compared to the state-of-the-art with code-based re-authentication?
		\item How does showing the authentication code inside the email subject line and body affect the authentication time compared to showing the authentication code only inside the email body?
	\end{RQ2LIST}
	\item \begin{RQ2LIST}
		\item Does the re-authentication method (e.g., code or link-based) affect the user behavior?
		\item Do the devices used for re-authentication (e.g., desktop or mobile) affect the user behavior?
	\end{RQ2LIST}
	\item How do users perceive different re-authentication methods?
\end{RQLIST}

\subsubsection{Contributions.}
We designed and conducted a between-group study with 592 participants recruited from the online service Mechanical Turk (MTurk)~\cite{kelley_conducting_2010} and evaluated the usability and perception of email-based re-authentication methods. Since there is still only one method used in practical deployments, we introduce two alternative RBA re-authentication methods, both of which have not yet been seen in the RBA context: a code-based and a link-based re-authenti\-cation scheme. We compared these approaches with the state-of-the-art RBA re-authentication method based on prior findings~\cite{wiefling_is_2019}.

Our results show that code-based methods have the potential to significantly speed up the re-authentication process while keeping the security properties at a similar level. We also identify significant differences in the perception of the re-authentication methods and provide recommendations.

Our work helps developers and website owners decide whether they should consider alternative re-authentication methods for RBA in their use case scenarios. Researchers obtain first insights on the perception of different email-based RBA re-authentication methods.

\section{Study}
\label{section:study}

To compare different RBA re-authentication methods, we designed a between-group usability study based on a specifically developed website. On this website, the participants registered a user account, providing a username and password as login credentials. %
After registering, participants were prompted to log in. When submitting the login credentials, the participants were asked for re-authentication through an email associated with the user account. Each participant perceived one of three different re-authentication methods, depending on the three study conditions below:

\begin{enumerate}
	\item \textbf{State of the Art} (SOTA): %
	The email had a six-digit authentication code in the body, which needed to be entered on the online service.
	\item \textbf{Subject} (SUBJ): %
	The email contained the authentication code in both subject line and body, which had to be entered on the online service.
	\item \textbf{Link} (LINK): %
	The email body contained an URL link, which had to be opened to confirm the authentication.
\end{enumerate}

We chose these re-authentication methods based on state-of-the-art RBA deployments~\cite{wiefling_is_2019} and email based verification methods known from popular online services. For the evaluation, we also subdivided the devices into three combinations, which we found realistic for practical RBA use case scenarios:

\begin{enumerate}
	\item \textbf{Desktop/Desktop}: The participants used a desktop PC on the website and also checked the email with this device.
	\item \textbf{Desktop/Mobile}: The participants used a desktop PC on the website and checked the email with a mobile device.
	\item \textbf{Mobile/Mobile}: The participants used a mobile device on the website and also checked the email with this device.
\end{enumerate}

We did not test Mobile/Desktop since we considered it to be an unrealistic use case scenario for RBA. We assume that most mobile devices have a pre-installed email app, making it unnecessary for users to check their email on a desktop PC while using a mobile device for the website.

\subsection{Design Decisions}
\begin{figure}[b]
	\vspace{-1em}
	\centering
	\subfloat[][SOTA, SUBJ]{
		\includegraphics[width=0.26\linewidth]{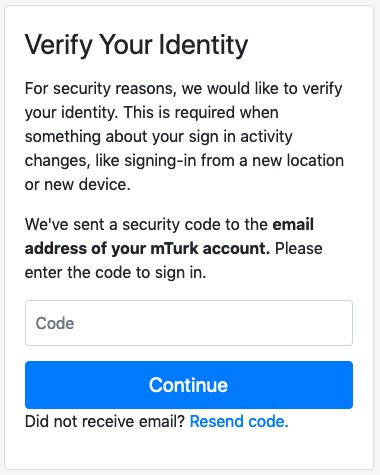}
		\label{fig:dialog-code}
	}
	\qquad
	\subfloat[][LINK]{
		\includegraphics[width=0.26\linewidth]{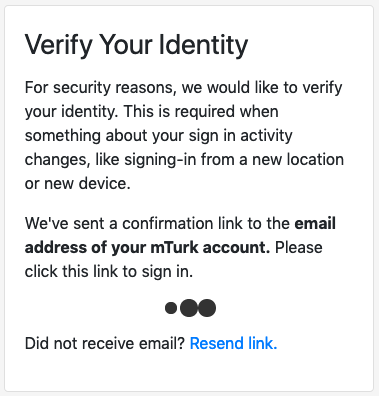}
		\label{fig:dialog-link}
	}
	\caption{Presented dialog types for the different study conditions}
	\label{fig:dialog}
\end{figure}

The dialogs and email contents for re-authentication differed in each condition. We outline the differences and design criteria in the following (see Figure~\ref{fig:dialog}).

\textbf{State of the Art (SOTA).} In previous work, we measured how RBA is used on popular online services~\cite{wiefling_is_2019}. We analyzed the Alexa Top 50 for RBA properties and extracted the RBA dialogs, if RBA was in use. %
Based on these observations on state-of-the-art RBA deployments, we designed a generic RBA dialog and confirmation email that we used in the study. We put text characteristics of dialogs and emails into categories and took the characteristics with the highest occurrences into the final dialog and email (see Figure~\ref{fig:dialog-code}).

\textbf{Subject (SUBJ).}
Authentication codes in the email subject line have been unknown in terms of RBA so far. However, we see potential in improving authentication speed and usability since the code is visible before opening the email, e.g., via push notifications on mobile devices. Codes in both subject line and email body are often used in email verification when registering a new user on a website. Both re-authentication dialog and email body are similar to those presented in SOTA. %
For the subject line, we collected account registration emails of popular online services that were using authentication codes in both subject line and body. Based on emails of LinkedIn, Facebook, and Slack, we created a generic subject line.

\textbf{Link (LINK).}
The link re-authentication method has not been seen in the context of RBA yet. We based this method on similar methods using a link for signing in (\emph{``magic links''}), used by the popular online services Tumblr, Medium, and Slack~\cite{van_amstel_should_2018}. %
\begin{figure}[t]
	\centering
	\subfloat[][Confirmation dialog]{
		\makebox[0.4\linewidth]{
			\includegraphics[width=0.26\linewidth]{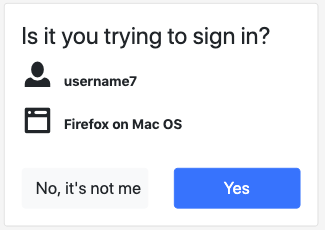}
			\label{fig:link-confirm-1}
		}
	}
	\hspace*{\fill}
	\subfloat[][After confirming the desktop device on a mobile device in the Desktop/Mobile scenario]{
		\makebox[0.53\linewidth]{
			\includegraphics[width=0.26\linewidth]{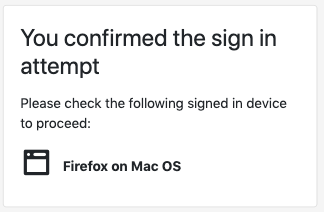}
			\label{fig:link-confirm-2}
		}
	}
	\caption{Re-authentication dialogs for the access confirmation in the LINK condition}
	\vspace{-1em}
\end{figure}
We adjusted the workflow for the RBA use case as follows: After entering the correct login credentials, the user received an email containing a link. The link contained a random verification string only known to the online service. We slightly changed the confirmation dialog to match the link confirmation use case (see Figure~\ref{fig:dialog-link}). When opening this link, the user was asked to confirm the device for signing in (see Figure~\ref{fig:link-confirm-1}). We based the dialog on Google's Android device confirmation dialog~\cite{google_sign_2019}.
After the user confirmed the device, this confirmed device was signed in. If the device that confirmed the login differed from the confirmed device, e.g., mobile device in the Desktop/Mobile use case, the user was advised to check the signed in device to proceed (see Figure~\ref{fig:link-confirm-2}). We did the additional confirmation to prevent that link prefetching via GET requests~\cite{komoroske_prerendering_2011} would cause the confirmation to be successful, i.e., we required an additional POST request to confirm the device. %
We tested this re-authentication method since the lack of entering a code has the potential to improve authentication speed and usability.

\subsection{Attacker Models}

In order to analyze our re-authentication methods in terms of usability metrics, their security properties have to be comparable with state-of-the-art deployments. Thus, we compare their online guessing security properties with three attacker models derived from known attacks on password-based authentication~\cite{freeman_who_2016}. We assume that the victim uses different passwords for the targeted online service and the email account. We also assume that the email provider blocks access to accounts after a number of wrong password entries (rate limiting). The attacker does not have physical access or eye contact with the victim's devices.%

The \textbf{password guesser} is a weak attacker that tries to guess the password of the victim, either by using brute-force or a list of popular passwords. When guessing the victim's password correctly, attackers still need to guess the email password, making the attack rather impractical.
Thus, this attacker will not be able to bypass all targeted re-authentication methods with reasonable effort.

The \textbf{credential stuffing attacker} is a rather strong attacker that has access to login credentials of the victim. The credentials are sourced from a password database leak of a different online service but are identical to the targeted one. Assuming that the password of the email account is not leaked, this attacker will not be able to bypass all targeted re-authentication methods.

The \textbf{phishing attacker} is a very strong attacker that tricks the victim to reveal the correct login credentials%
. The attacker sets up a website on a phishing domain imitating the appearance of the targeted online service. The degree of imitation varies from simply copying the HTML code of the targeted online service to forwarding the complete traffic between victim and online service (man in the middle, MITM). On success, attackers obtain the victim's login credentials. For MITM, attackers can even forward the entered authentication code to the online service, bypassing the re-authentication%
. However, attackers cannot bypass email verification links, since the phishing domain is not included in the email verification link. Thus, the link verification is conducted at the real online service%
. Assuming that the email password is not leaked, a phishing attacker could bypass SOTA and SUBJ but not LINK.

\subsection{Study Design}

We decided to conduct a two-part between-group study to compare different re-authentication methods of RBA in terms of authentication time and user perception, and to measure the behavior when perceiving this re-authentication on the website for the first time. The study consisted of two parts:%

\vspace{-1em}
\subsubsection{Login.}

First, the participants registered on the study website with username and password. The website was reachable via HTTPS via an internet domain not linked to our university to mitigate social desirability bias~\cite{nederhof_methods_1985}. After registering, the participants tried logging into the website. After submitting the correct login credentials, the website asked for re-authentication, which differed between the three conditions SOTA, SUBJ, and LINK.

\vspace{-1em}
\subsubsection{Exit Survey.}

After completing the re-authentication, the participants answered a short survey. The questions were presented in random order to randomly distribute ordering effects~\cite{kalton_effect_1982}. The order of response options were also randomized in each question to randomly distribute response order bias~\cite{chan_response-order_1991,hartley_thoughts_2014}.

In the survey, the participants stated in a free text answer the device on which they opened the identity verification email on, %
to determine if the device used for verification is a desktop or a mobile device. They also listed in free text answers three feelings they had when they were asked to verify their identity. This question was inspired by Golla et al.~\cite{golla_what_2018}. We used it to discover the user perceptions of the re-authentication. The participants also answered, by ticking checkboxes, which online services they used in the last month. %
The list of online services included the response option \emph{MTurk} as an attention check to verify the quality of our results~\cite{abbey_attention_2017}. 
The survey concluded with %
demographic questions.

\subsection{Data Collection}

To answer our research questions, we collected the following data:
\begin{enumerate*}
	\item \textbf{Timing and event information}: We collected timestamps of when certain events occurred on the website%
	. %
	We used the timestamps to calculate durations for parts of the re-authentication process%
	.
	In addition, we used the recorded events to analyze the participants' behavior during re-authentication%
	.
	\item \textbf{Device information}: We collected the user agent string of the device that the participant used to log in on the website. On the LINK condition, we also collected the user agent string of the device that opened the verification link. We used this information to determine the devices as mobile or desktop devices. We also used this information to verify in the LINK condition if the survey answer regarding the used device was correct. This enabled us to increase the quality of the collected data.%
	\item \textbf{Survey answers}: We stored the survey responses digitally and analyzed them after the study.
\end{enumerate*}

\subsection{Data Processing}
\label{subsection:data-processing}

After collecting the data, we processed the data as follows:

\subsubsection{Devices.}

We subdivided our data set into the three different device combinations Desktop/Desktop, Desktop/Mobile, and Mobile/Mobile. We determined the device used for logging in with the recorded user agent string.
Due to the different properties of the code and link-based conditions, we determined the email checking device as follows. For the code based conditions SOTA and SUBJ, we checked the corresponding free text responses given by the participants and classified them into the categories mobile or desktop device. %
For LINK, we also checked the user agent string of the device that clicked the link%
.

In the Desktop/Mobile use case, we furthermore analyzed the recorded browser events to verify the given answer of the participant. If the event log showed that the participant copied and pasted the code, which is not possible for all setups except for those using the macOS Universal Clipboard feature%
, we assumed that the participant gave an invalid response and filtered this response.%

\subsubsection{Times.}

We calculated different types of times from the timestamp information. We measured the times to find out whether one of the re-authentication methods is completed faster in parts of the re-authentication process than the other.

\begin{enumerate*}
	\item \textbf{Challenge Completion Time}:
	We measured the time needed to complete the re-authentication challenge. In %
	the code-based challenges (SOTA and SUBJ), the time was calculated as the timestamp differences between submitting the code and the last focus event before entering the code. We decided to take the last focus event since we needed to consider the delay between understanding the user interface and conducting the code entering action. Also, when opening the link in the LINK condition, the window is focused in that moment as well, making LINK comparable to SOTA and SUBJ. In the Desktop/Mobile case, %
	we took the timestamp differences between submitting the code and the beginning of the code entering. Though we took a different timestamp in this case, we expect the overall time for Desktop/Mobile to be higher than for Desktop/Desktop and Mobile/Mobile anyway since the code has to be entered manually. By doing this, we aimed to ensure comparability between the code and link-based re-authentication methods in any use case scenario.
	
	\item \textbf{Re-Authentication Duration}: We also measured the time %
	needed for the re-authentication in total. We calculated this time as the difference between finishing the  re-authentication challenge and loading the identity confirmation dialog for the first time.
\end{enumerate*}

\subsubsection{Feelings.}

From the feelings provided in the open ended question, we corrected the grammar, and converted nouns and verbs to adjectives with the WordNet~\cite{miller_wordnet:_1995} database where applicable. We did this to correct misspellings and differences in tenses. We also clustered the feelings with Emolex~\cite{mohammad_crowdsourcing_2013} into the categories positive, neutral, and negative, to analyze the sentiment towards the perceived re-authentication method. This approach was similar to %
Golla et al.~\cite{golla_what_2018}.

\subsection{Piloting}

We did a pilot study with 10 participants to test and verify our study procedure. After the pilot study, we added additional measurements and slightly changed some dialogs on the website as a result of piloting. Participants involved in the pilot study were excluded from the final study to avoid bias.

\subsection{Recruiting}

We recruited participants via the crowdworker platform MTurk, which has shown to be applicable for usability studies involving short reactional tasks~\cite{kelley_conducting_2010}. %
We required the participants to %
be 18 years or older, and have a 95\% task approval rate. The study was advertised as a website testing study that is expected to take 10~minutes. We did not mention that we test authentication schemes to avoid bias. Each participant was compensated with \$1.64 after study completion.%

Each participant was randomly assigned to one of the three conditions while keeping the group size of each condition as equal as possible.%

\subsection{Ethical Considerations}

We made sure to meet the needs of the MTurk participants (clickworkers) for ethical issues and to improve our data quality. We offered the clickworkers more flexibility by increasing the task time to 24 hours since it has shown to both speeding up task completion and improving the result quality~\cite{yin_running_2018}. Rejected work on MTurk can result in clickworkers losing qualifications on the platform, affecting their monthly income. Thus, we communicated to the workers that we do not reject any work to make them feel comfortable~\cite{hara_data-driven_2018}. We followed the paying recommendations by Hara et al.~\cite{hara_data-driven_2018}, having in mind that workers are not paid between MTurk tasks. In order for the clickworkers to make a living, we set the compensation so high that it is possible for them to earn more than the hourly minimum wage of their home country, i.e., \$7.25/hr in the US.
We did not collect any email addresses, as this is against MTurk's acceptable use policy. Instead, the MTurk service sent the emails out to the participants.
All participants %
gave informed consent. All questions offered a ``don't know'' option.

We do not have a formal IRB process at TH Köln, where we conducted this study, but besides our ethical considerations above, we made sure to minimize potential harm by complying with the ethics code of the German Sociological Association (DGS) as well as the standards of good scientific practice of the German Research Foundation (DFG). We also made sure to comply with the terms of the EU General Data Protection Regulation.

\section{Results}
\label{section:results}

The study took place between July and October 2019 and a total of 592 users participated. 499 participants completed the study. From these participants, 48 were excluded from the set for the following reasons: 
\begin{enumerate*}
	\item They copied and pasted the authentication code while stating that they used a specific Desktop/Mobile setup in which this is technically not feasible (n=19).
	\item They failed the attention check (n=13).
	\item They used a mobile device on the website and checked the email with a desktop PC, which we did not test in our study (n=11).
	\item We were unable to determine the device based on the participant's free text answer %
	(n=5).
\end{enumerate*}
The dropouts were similarly distributed across all conditions.

At the end, we retained 451 participants for the analysis. Table~\ref{tab:participants} shows how these were distributed among the different conditions and device combinations. %
The participants completed the study in four minutes on median average.

\setlength\tabcolsep{4pt}
\begin{table}[t]
	\centering
	\caption{Number of participants in each condition and device use case scenario}
		\begin{tabular}{@{}llll@{}}
			\toprule
			Website/Email & SOTA & SUBJ & LINK \\ 
			\midrule
			Desktop/Desktop & 67 & 67 & 72 \\ 
			
			Desktop/Mobile & 50 & 45 & 48 \\ 
			
			Mobile/Mobile & 30 & 36 & 36 \\ 
			\bottomrule
		\end{tabular}%
	\label{tab:participants}
\end{table}

Our participants were 53.6\% female, 45.0\% male, and 0.2\% non-binary. The age of the participants ranged from 18 to 74. The majority of participants were between 25 and 34 years old (41.9\%), while 11.3\% were younger and 46.4\% were older. The remaining percentages preferred not to answer the corresponding demographical question. The majority of participants had an associate degree or higher (62.8\%) and did not have a computer science background (75.4\%).

For statistical analysis of the timing data, we used Kruskal-Wallis %
tests for the omnibus case and Dunn's multiple comparison test with Bonferroni correction %
for post-hoc analysis. For categorical data, i.e., the feelings and number of login attempts, we used Pearson's chi-square test for contingency table analysis ($\chi ^2$). We set 0.05 as the threshold for statistical significance, i.e., p~\textless~0.05 is significant.
In the following, we outline the results ordered by the research questions given in Section~\ref{subsection:research-questions}. A discussion follows after the results of each research question.

\subsection{Authentication Times (RQ1)}

\subsubsection{Challenge Completion Time.}
\label{subsubsection:challenge-completion-time}

\begin{figure}[t]
	\vspace{-1em}
	\centering
	\subfloat[][Desktop/Desktop]{
	\includegraphics[width=0.31\linewidth]{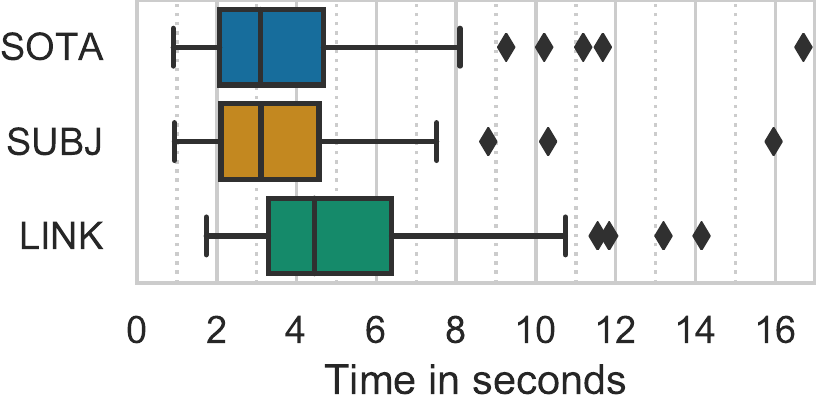}
	\label{fig:reactionduration-desktop-desktop}
	}
	\hspace*{\fill}
	\subfloat[][Desktop/Mobile]{
	\includegraphics[width=0.31\linewidth]{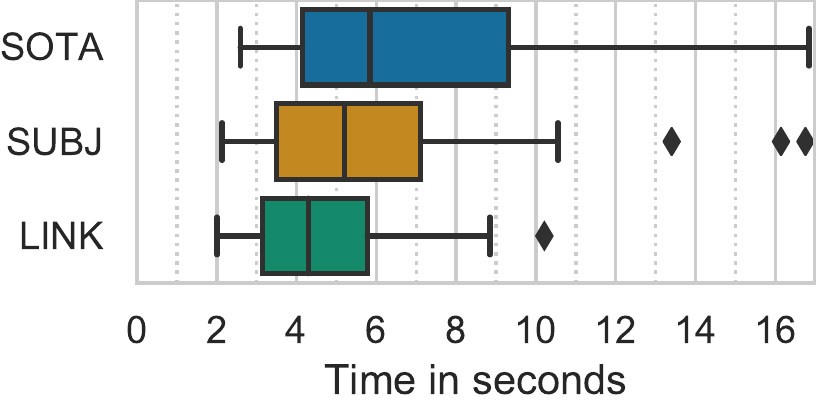}
	\label{fig:reactionduration-desktop-mobile}
	}
	\hspace*{\fill}
	\subfloat[][Mobile/Mobile]{
			\includegraphics[width=0.31\linewidth]{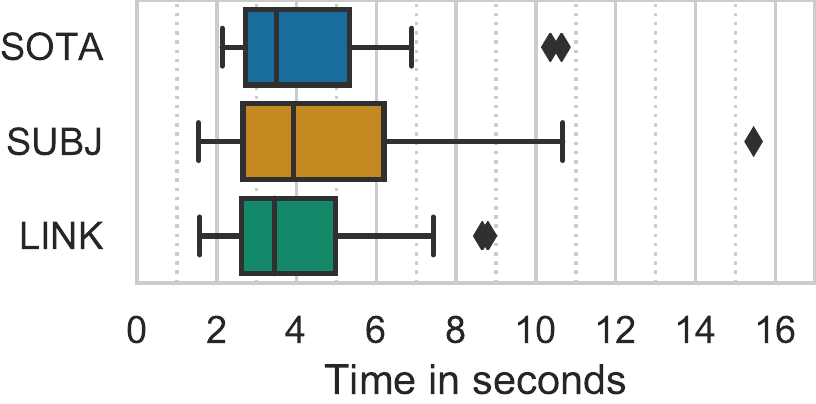}
			\label{fig:reactionduration-mobile-mobile}
		}
	\caption{Challenge completion times for the conditions and device combinations. There are significant differences in Desktop/Desktop and Desktop/Mobile.}
	\label{fig:reactionduration}
\end{figure}

The participants completed the re-authentication challenge with median times between three and six seconds (see Figure~\ref{fig:reactionduration}). There were significant differences in some conditions and device combinations.

For Desktop/Desktop, the challenge completion time for LINK was significantly higher than those for SOTA and SUBJ (LINK\slash SOTA: p=0.0024; LINK\slash SUBJ: p=0.0009). For Desktop/Mobile, the challenge completion time for LINK was significantly lower than for SOTA (p=0.0038). For Mobile/Mobile, there were no significant differences between all three conditions.

Completing the re-authentication challenge took significantly more time on Desktop/Mobile than on Desktop/Desktop for the code-based conditions (SOTA: p$<$0.0001; SUBJ: p=0.0002). For SOTA in addition, challenge completion took significantly more time on Desktop/Mobile than on Mobile/Mobile (p=0.0069).

Concluding the results, link-based authentication challenges were solved faster than the code-based ones when they were not solved on the same device that they used for the login attempt. In the other cases, they were either solved slower (Desktop/Desktop) or with similar speed (Mobile/Mobile). Showing the authentication code inside the email subject did not have a significant effect on the challenge completion time.

\textbf{Discussion:} In contrast to SOTA and SUBJ, LINK participants had to check their device in an extra confirmation dialog and therefore loaded an additional web page, which is why we assume that they needed more time on Desktop/Desktop to complete the challenge.
Since all participants on Desktop/Mobile could only manually enter the code, this explains the increased challenge completion time for the code-based challenges on this device combination.

\vspace{-1em}
\subsubsection{Re-Authentication Duration.}

\begin{figure}[t]
	\vspace{-1.2em}
	\centering
	\subfloat[][Desktop/Desktop]{
	\includegraphics[width=0.31\linewidth]{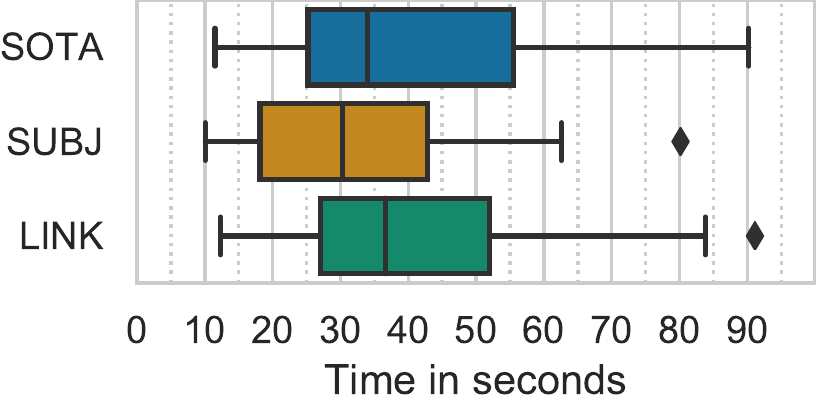}
	\label{fig:loginduration-desktop-desktop}
	}
	\hspace*{\fill}
	\subfloat[][Desktop/Mobile]{
	\includegraphics[width=0.31\linewidth]{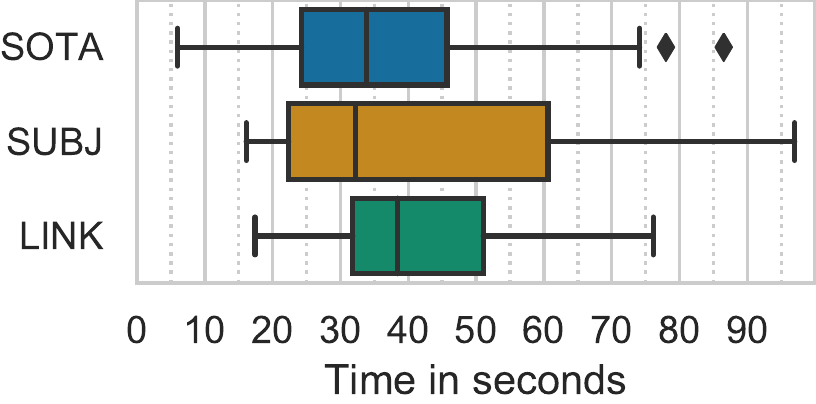}
	\label{fig:loginduration-desktop-mobile}
	}
	\hspace*{\fill}
	\subfloat[][Mobile/Mobile]{
			\includegraphics[width=0.31\linewidth]{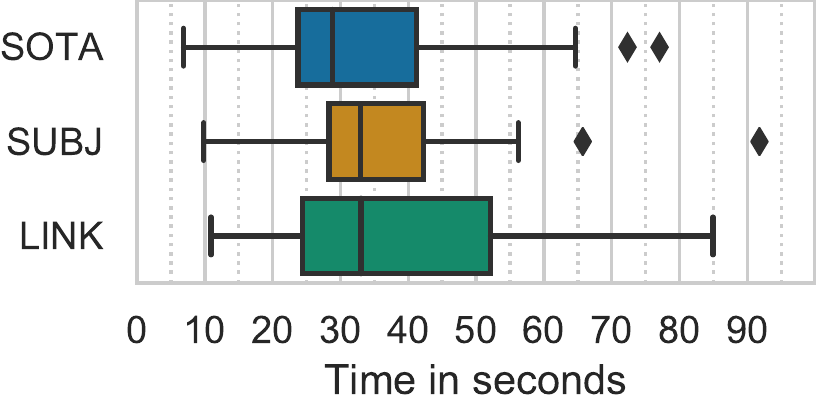}
			\label{fig:loginduration-mobile-mobile}
		}
	\caption{Re-authentication duration for the conditions and device combinations. The difference between LINK and SUBJ in Desktop/Desktop is significant.%
	}
	\label{fig:loginduration}
	\vspace{-1em}
\end{figure}

In summary for all participants, it took a median of 33.82 seconds to re-authenticate (mean: 71.89s, std: 398.22s). %
For the Desktop/Desktop combination (see Figure~\ref{fig:loginduration-desktop-desktop}), the overall re-authentication time for SUBJ was significantly lower than for LINK (p=0.0226). For all the other conditions, we could not find any significant differences.%

Concluding the results, showing the authentication code inside the email subject decreased the re-authentication time compared to link-based authentication. However, it did not significantly affect the re-authentication time compared to showing the authentication code only inside the email body. Also, link-based authentication did not significantly affect the authentication time compared to the state-of-the-art code-based authentication.

\textbf{Discussion:} Since there were significant differences, we assume that showing the code in the subject line affected the login duration in total. %
Opening a link introduces a delay to load the target website. Some email providers also introduce additional delays when clicking on a link, mostly to advise their users that they are redirected to another website. As a result, participants using login links will always experience a constant delay. This explains the significantly longer login duration for LINK. We assume that the faster login duration for SUBJ with Desktop\slash Desktop combination lies in the fact that the participants saw the authentication code earlier and thus did not have to open the email to receive it. In summary, the email delivery and opening is the biggest factor affecting the login duration. Thus, we suggest that this email based re-authentication should not be asked too often, which is the case with RBA.

\subsection{Behavior During Authentication (RQ2)}
\label{subsection:behavior}

Most SUBJ and SOTA users and all LINK users passed the re-authentication challenge on the first attempt (SOTA: 95.2\%, SUBJ: 98.0\%, LINK: 100\%). The remaining participants passed the challenge on the second attempt. %

The majority of participants in the code-based conditions copied and pasted the code into the code entering form when the device combination allowed it (Desktop/Desktop: 88.1\%; Mobile/Mobile: 59.1\%).
Concluding these results, code-based re-authentication schemes have the tendency to cause users to copy and paste the code when conducted on the same device.

\textbf{Discussion:} %
We assume that copying and pasting the code was the main reason why the code-based challenges were solved faster than the link-based challenges when solved on the same device%
. Our results reflect findings of Doerfler et al.~\cite{doerfler_evaluating_2019} regarding a high success rate for email-based re-authentication.

\subsection{Perceptions (RQ3)}

\begin{figure}[t]
	\vspace{-1em}
	\centering
	\subfloat[][SOTA]{
		\includegraphics[width=0.327\linewidth]{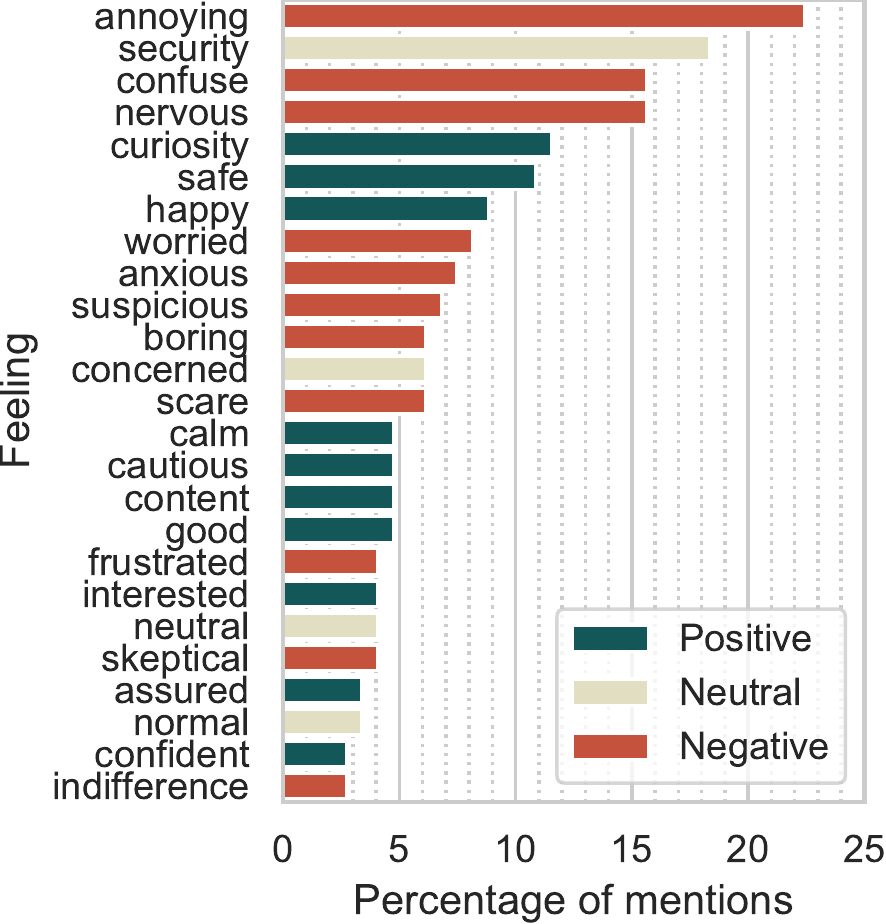}
		\label{fig:emotions-2fa-sota}
	}
	\hspace*{\fill}
	\subfloat[][SUBJ]{
		\includegraphics[width=0.32\linewidth]{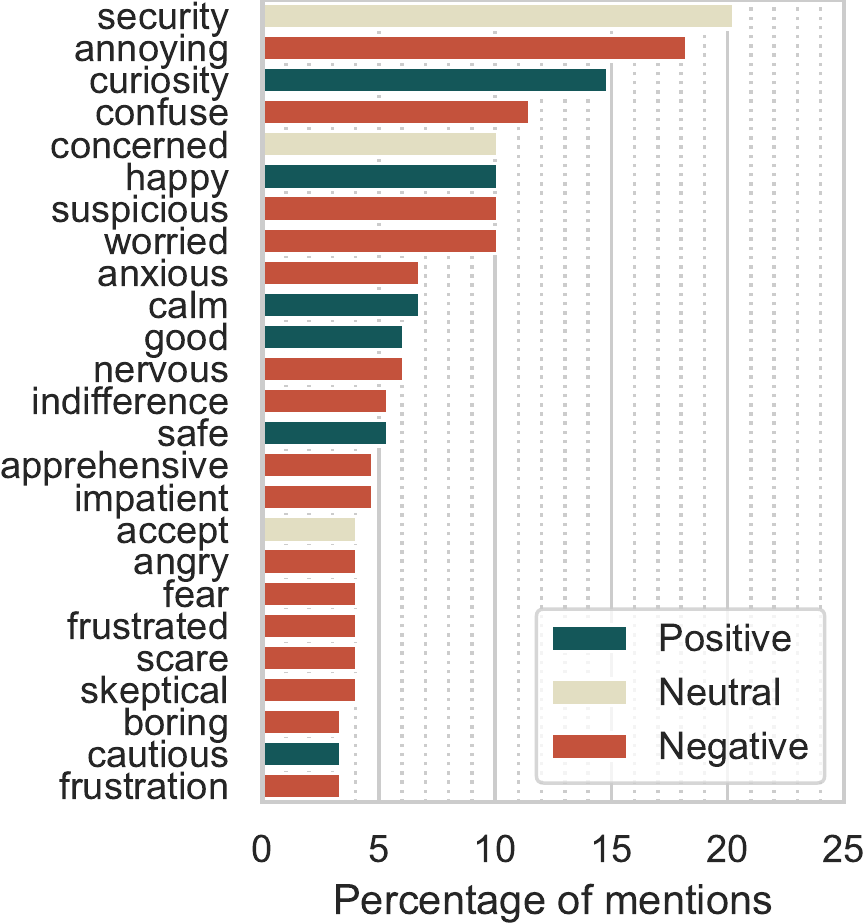}
		\label{fig:emotions-2fa-subj}
	}
	\hspace*{\fill}
	\subfloat[][LINK]{
		\includegraphics[width=0.31\linewidth]{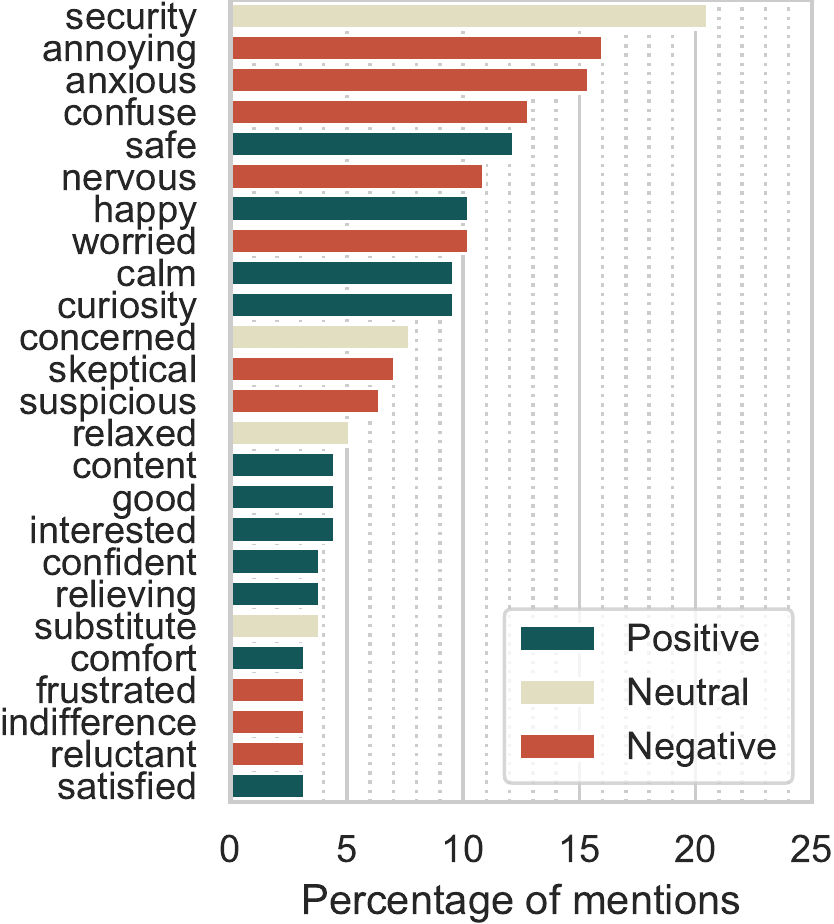}
		\label{fig:emotions-2fa-link}
	}
	\caption{Feelings the participants had when asked to verify their identity}
	\label{fig:emotions}
	\vspace{-1.05em}
\end{figure}

All participants listed three feelings they had after they were asked to verify their identity. Figure~\ref{fig:emotions} shows the 25 most mentioned feelings ordered by the number of occurrences. The re-authentication methods resulted in mixed emotions. While there was no clear tendency for positive or negative feelings in SOTA and LINK, the top 25 feelings in SUBJ were more negative. The feelings \emph{security} and \emph{annoying} were the most mentioned ones in all three conditions. We discovered significant differences between the three conditions for anxious, nervous and neutral (see Table~\ref{tab:feelings-significance}).%
\setlength\tabcolsep{4pt}
\begin{table}[b]
	\vspace{-1em}
	\centering
	\caption{Significant $\chi ^2$ results for the mentioned feelings in each condition and the percentage of mentions in each condition.}
		\begin{tabular}{@{}l|rr|rrr@{}}
			\toprule
			Feeling &   $\chi ^2$ &       p & SOTA & SUBJ & LINK \\
			\midrule
			anxious &  7.8053 &  0.0202 &         7.5\% &         6.8\% &        15.4\% \\
			nervous &  6.9677 &  0.0307 &        15.6\% &         6.1\% &        10.9\% \\
			neutral &  6.6667 &  0.0357 &         4.1\% &         0.7\% &         0.6\% \\
			\bottomrule
		\end{tabular}%
	\label{tab:feelings-significance}
\end{table}%
The other feelings were mentioned in similar occurrences across all categories. The most mentioned positive feelings were curiosity, happy, safe, calm, and good. For the neutral direction, these were security, concerned, relaxed, substitute, and accept. The most mentioned negative feelings were annoying, confuse, nervous, anxious, and worried.

\textbf{Discussion:} Due to phishing awareness campaigns and trainings, users are trained not to open links in emails~\cite{sheng_who_2010}. Being asked to click on a link in an email for authentication contradicts the trained behavior, resulting in an insecure feeling. We assume that this explains why participants named the anxious feeling significantly more often in LINK%
. However, it is possible that this anxious feeling declines when repeating the link-based re-authentication procedure multiple times~\cite{zajonc_attitudinal_1968}. There are differences between re-authentication emails and phishing emails that support this assumption. First, the website accessed by the link does not require login credentials. Second, we assume that users expect this re-authentication email to appear in their email inbox shortly%
.

SUBJ participants did not need to open the email to get the authentication code. %
SOTA and LINK participants had to open an email whose contents they had never seen before, i.e., the code or link. We assume that this is why SUBJ participants named a nervous feeling less often than those of SOTA and LINK.

\section{Limitations}
\label{section:limitations}

The results are limited to a part of a population of a specific country. We assume that the self-reported answers were typical for participants from the US with college education that are younger than 50 years~\cite{redmiles_how_2019}. Due to the restrictions of MTurk, we could only test email address verification for plain text emails. It is possible that HTML emails are perceived differently by participants~\cite{karakasiliotis_assessing_2006}.%

Since the participants were only authenticating once, we assume that they expected the re-authentication for every login attempt when reporting the feelings. Following that, we assume that the results were more related to 2FA than for RBA. Since users tend to disable re-authentication when asked too often~\cite{crawford_understanding_2014}, we assume that the feelings results would be more positive in the real world. 

\section{Related Work}
\label{section:related-work}
RBA re-authentication challenges were not evaluated in literature so far. There are related studies evaluating other authentication methods.
De Cristofaro et al.~\cite{de_cristofaro_comparative_2014} compared three 2FA solutions with a study involving MTurk participants. In contrast to our study, their participants were not exposed to RBA solutions.
Agarwal et al.~\cite{agarwal_ask_2016} evaluated four re-authentication methods for smartphones. Similar to our study, they introduced new re-authentication methods %
and exposed their participants to them. However, these re-authentication methods were only applicable for mobile apps and thus were not suitable for RBA in general.

Doerfler et al.~\cite{doerfler_evaluating_2019} evaluated the effectiveness of Google's re-authentication challenges by analyzing login attempt data. Their results showed that code-based re-authentication protected against more than 90\% of all phishing attempts. Although this shows the effectiveness of RBA against phishing, no usability metrics are examined in their work that study its characteristic and potentials.

\section{Conclusion}
\label{section:conclusion}
As long as online services continue to use password-based authentication, RBA is becoming increasingly important as a complementary protection measure. This is further underlined by the fact that RBA is explicitly recommended by NIST~\cite{grassi_digital_2017}. However, there is little scientific research focused on RBA so far. Its development is mainly driven by online services that already use RBA. Since these are popular online services, they have a major impact on the state-of-the-art deployment as can be derived from the single re-authentication method. No scientific evaluation indicates that this is the most appropriate approach to use for implementation.

Our study closes this gap and compares the state-of-the-art email-based RBA re-authentication method with two introduced alternatives regarding their time exposure, security, and user-perceived security. Our results indicate that link-based re-authentication results in higher time requirements and anxiety when perceived for the first time. Code-based re-authentication has proven to be more advantageous in this respect. More specifically, showing the authentication code in the subject line has the potential to reduce re-authentication time with perceptions comparable to the state-of-the-art deployment. Following that, website owners should carefully adjust their RBA re-authentication design to be appropriate for their applications. In general, our research suggests that further research should study RBA more consistently so that all services can benefit from reliable scientific results while hardening password authentication with RBA.

\vspace{-1em}
\subsubsection*{Acknowledgments.}

This research was supported by %
NERD.NRW %
sponsored by the state of North Rhine-Westphalia. The research was also supported by a RISE Germany scholarship granted by the German Academic Exchange Service (DAAD) and sponsored by the German Federal Foreign Office.

	\appendix

	 \bibliographystyle{splncs04}
	 \bibliography{bibliography}
\end{document}